\documentclass[fleqn,usenatbib]{mnras}

\usepackage{mathptmx}

\usepackage[T1]{fontenc}

\DeclareRobustCommand{\VAN}[3]{#2}
\let\VANthebibliography\thebibliography
\def\thebibliography{\DeclareRobustCommand{\VAN}[3]{##3}\VANthebibliography}

\usepackage{graphicx}   
\usepackage{amsmath}    
\usepackage{amssymb}    

\title[UHECRs from Andromeda galaxy]{Ultra high energy cosmic rays from past activity of Andromeda galaxy}

\author[V. N. Zirakashvili et al.]{
V. N. Zirakashvili,\thanks{E-mail: zirak@izmiran.ru} V. S. Ptuskin, and S.I.Rogovaya
\\
Pushkov Institute of Terrestrial Magnetism, Ionosphere and Radiowave
Propagation, 108840, Troitsk, Moscow, Russia}

\date{Accepted 2022 November 8. Received 2022 October 25; in original form 2022 September 21}

\pubyear{2022}

\begin{document}
\label{firstpage}
\pagerange{\pageref{firstpage}--\pageref{lastpage}}
\maketitle

\begin{abstract}
It is shown that the relativistic jets associated with the growth and past activity
of the supermassive black hole in the Andromeda galaxy could be the main source
 of cosmic rays with energies above $10^{15}$ eV. Most of the cosmic ray energy is related to a bow shock of the
 jet that produces multi-PeV cosmic rays with light composition.
The highest energy cosmic rays with heavy composition are produced in the jet itself.
The spectra of energetic particles produced in Andromeda galaxy and propagated to the Earth
are calculated and
compared with observations.
\end{abstract}




\section{Introduction}

It is believed that the acceleration of ultra-high energy cosmic rays (UHECRs) occurs
 in astrophysical objects with  relativistic motions. This is happening in pulsar nebula,
 active galactic nuclei (AGN) jets driven by the gas accretion onto central black holes, and anisotropic
explosions of gamma-ray bursts (see a review of \citet{bykov12}).

It is  currently clear, that the energy released during the gas accretion and the corresponding growth of
 supermassive black holes (SMBHs) produce a strong impact on the evolution of parent galaxies
(e.g. \citet{donahue22}).
If the power of the jet driven by accretion is high enough the jet can  propagate on sub-Mpc scales in the
circumgalactic medium. The particles can be accelerated in the magnetosphere of rotating SMBH
\citep{istomin09, banados09, jacobson10, wei10}, near the jet
 boundary via shear acceleration, at the inner termination shock at the end of the jet,
and the outer bow shock surrounding the cocoon of the jet \citep{norman95}.

Taking into account different acceleration sites and different acceleration mechanisms one
can expect several different components in the general flux
 of emitted accelerated particles.

In particular, the spectra of energetic particles produced at the outer bow
shock of the jet via the diffusive shock acceleration (DSA) mechanism
are similar to the spectra of galactic supernova remnants (SNRs). It
is known that the DSA mechanism
 \citep{krymsky77,bell78,axford77,blandford78} operates
 in the vicinity of shocks in SNRs. The X-ray and gamma-ray observations
of the last decades indicated the presence of multi-TeV energetic
particles in these objects (see e.g. \citet{lemoine14} for a
review).

The light chemical composition of cosmic rays accelerated at the bow
shock is expected because of the low metallicity and high ionization
state of
 the circumgalactic medium.

The heavier composition and different spectra are expected for
particles accelerated in the jet itself. Acceleration of particles
in the shear flow \citep{berezhko81, earl88}
 is one of the possibilities. The presence of electrons accelerated up
to  PeV energies in shear flows is consistent with modern X-ray
and gamma-ray observations of large-scale extragalactic jets
\citep{wang21}. It is expected that protons and nuclei can be
accelerated to
 higher energies because of the lower energy losses.

Probably the jets with strong toroidal magnetic fields produce the highest energy particles
 while the maximum energy of particles accelerated by bow shocks is lower. Low-energy
 particles can propagate only from nearby sources. The prime candidates are SMBHs in the Galactic center and
 the Andromeda galaxy with masses
$4\cdot 10^6\ M_{\odot}$ and $2\cdot 10^8\ M_{\odot}$ respectively. They are not in an active state now.
However, they were active in past. The
 huge gamma ray halo of Andromeda galaxy \citep{karwin19},  Fermi and eROSITA bubbles  \citep{su10,predehl20}
in the Milky Way are apparently originated as a result of recent
 activity of the central SMBHs. Cosmological simulations of Andromeda-like galaxies
\citep{pillepich21} also
 demonstrate a pulsed activity of SMBH every $10^8$ years
and the peak mechanical luminosity about $10^{44}$ erg s$^{-1}$.

The idea of the cosmic ray production during the past activity of
the Galactic center is not new (see e.g. \citet{ptuskin81, giler83,
istomin14, fujita17}). The particles above the "knee" energies could
be also produced in other objects like Galactic winds (e.g.
\citet{zirakashvili04}), peculiar supernovae (e.g. \citet{wang07}),
and neutron star mergers \citep{kimura18a}.

In the present paper we calculate the propagation of UHECRs from Galactic
center and nearby Andromeda galaxy and check whether they can considerably
contribute to the observed spectrum of UHECRs.

The paper is organized as follows. In the next Sections 2 and 3, we describe our model.
The application of the model for Andromeda  and Milky Way is given in Section 4.
The discussion of results and conclusions are presented in Sections 5 and 6.

\section{Propagation model}

The evolution of energy distributions of protons $N({\bf r},z,\epsilon )$ and nuclei $N_i({\bf r},z,\epsilon ,A)$
in expanding Universe is described by equations \citep{berezinsky06}

\begin{eqnarray}
-H(z)(z+1)\frac {\partial N}{\partial z}=\nabla D({\bf r},z,\epsilon )(z+1)^2 \nabla N
+H(z)\left( \epsilon \frac {\partial N}{\partial \epsilon}-2N\right)  \nonumber \\
+\frac {\partial }{\partial \epsilon }b(\epsilon )N
+4\nu_{ph}(4)N_i(4) +\sum _{A=5}^{56}\nu
_{ph}(A)N_i(A)+q(z,\epsilon ) (1+z)^3\delta ({\bf r}),
\end{eqnarray}

\begin{eqnarray}
-H(z)(z+1)\frac {\partial N_i(A)}{\partial z}=\nabla D_i({\bf r},z,\epsilon )(z+1)^2\nabla N_i(A) \nonumber \\
+H(z)\left( \epsilon \frac {\partial N_i(A)}{\partial \epsilon }-2N_i(A)\right)
+\frac {\partial }{\partial \epsilon}b(\epsilon)N_i(A) \nonumber \\
-\nu _{ph}(A)N_i(A)+\nu _{ph}(A+1)N_i(A+1)+q_i({z,\epsilon ,A)
(1+z)^3\delta (\bf r}).
\end{eqnarray}

Here ${\bf r}$ is the comoving coordinate, and the redshift $z$ is
used instead of time. This system for all kinds of nuclei with
different mass numbers $A$ from Iron to Hydrogen should be solved
simultaneously. The energy per nucleon $\epsilon=E/A$ is used here
because it is approximately conserved in a process of nuclear
photodisintegration, $q(z,\epsilon )$ and $q_i(z,\epsilon ,A)$ are
the spectra of the point cosmic-ray proton and nuclei sources
respectively, $b(A,\epsilon,z)$ is the characteristic rate of energy
loss by the production of $e^{-}e^{+}$ pairs and pions, $\nu
_{ph}(A,\epsilon,z)$ is the frequency of nuclear photodisintegration
(for details see our paper \citet{ptuskin13}), the sum in the right
side of Eq. (1) describes the contribution of secondary protons
produced by the photodisintegration of heavier nuclei,
$H(z)=H_{0}((1+z)^{3}\Omega_{m}+\Omega_{\Lambda})^{1/2}$ is the
Hubble parameter in a flat universe with the matter density
$\Omega_{m}(=0.3)$ and the $\Lambda$-term $\Omega_{\Lambda}(=0.7)$,
its value $H_0=70$ km s$^{-1}$ Mpc$^{-1}$ at current epoch is used.

Diffusion coefficient $D({\bf r},z,E)$ is determined by scattering on
magnetic inhomogeneities. We use the analytical approximation
obtained in numerical trajectory calculations of \citet{harari14}.

\begin{equation}
D= \frac {cl_c}{3}\left( 4\frac{E^2}{E^2_c}+0.9\frac{E}{E_c}
+0.23\frac{E^{1/3}}{E^{1/3}_c}\right) , \ E_c=ZeBl_c
\end{equation}
where $B\propto (1+z)^2$ is the magnetic field strength and $l_c\propto (1+z)^{-1}$
is the coherence scale of the magnetic field in the intergalactic medium.

\section{Particle acceleration in jets}

\begin{figure}
\begin{center}
\includegraphics[width=8.0cm]{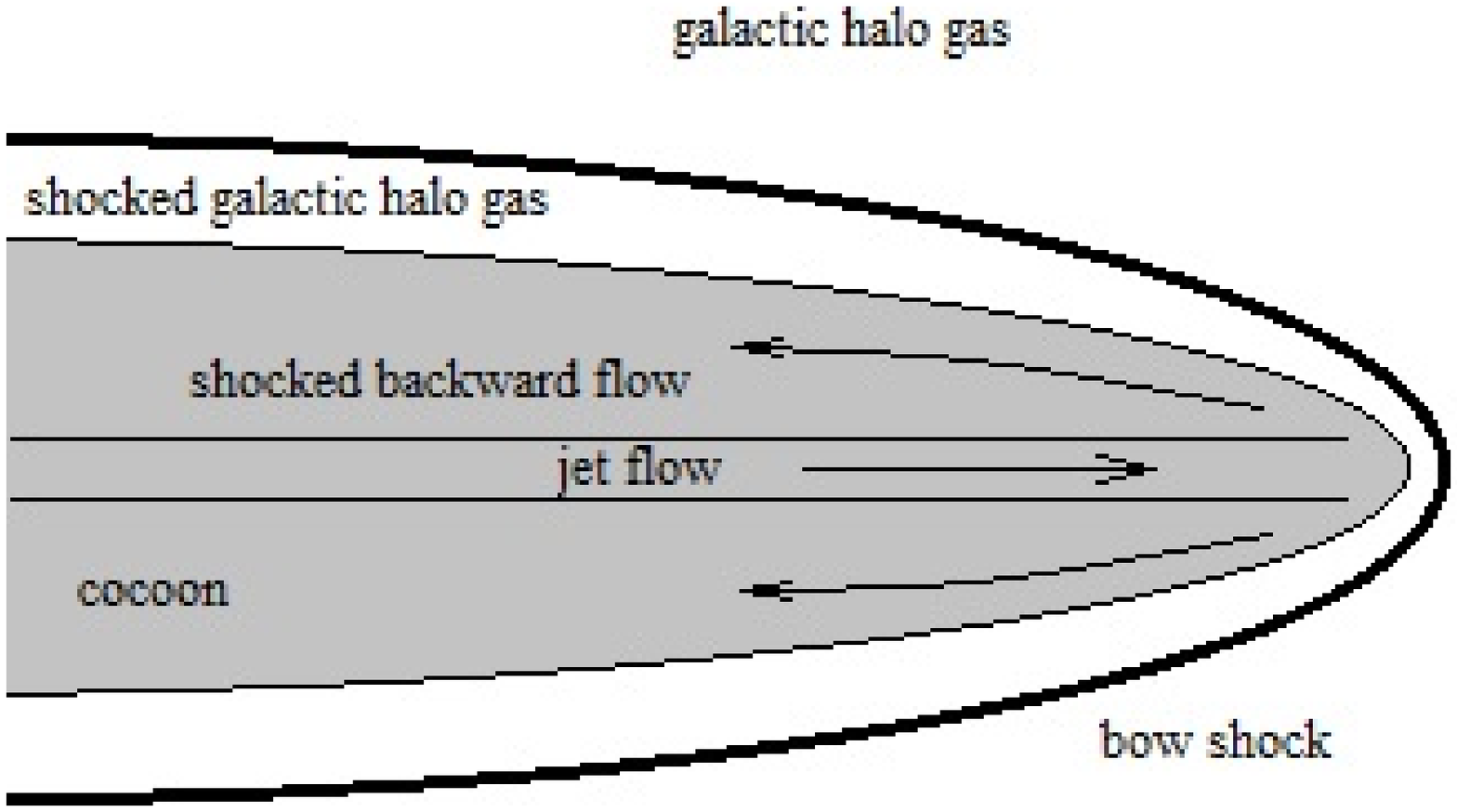}
\end{center}
\caption{ Schematical view of the jet. }
\end{figure}

A relativistic jet outflow produces complex flow structures in the
galactic halo schematically shown in Fig.1 (see, for example,
modeling of \citet{seo21}) . The supersonic jet flow terminates at
the end of the jet and produces a low density cocoon with the
backward flow. The cocoon is surrounded by the denser galactic halo
gas shocked at the bow shock. The bow shock propagates at a
non-relativistic speed depending on the ratio of the jet and halo
gas densities. DSA mechanism can operate at the bow shock,
 at the termination shock at the end of the jet, and in multiple small scale shocks inside the bow shock and cocoon observed
 in numerical modeling of jets \citep{seo21}. The presence of the shear flow in the
cocoon and in the jet itself probably results in the shear acceleration. Below we shall consider 3 main components of
accelerated particles.

1) The particles accelerated up to maximum energies  in the jet itself via shear acceleration, DSA at the termination shock,
 or instabilities in the jet.

2) The particles with the lowest maximum energies are accelerated at the non-relativistic bow shock.
The properties of this component are robust because DSA is well studied.

3) The particles accelerated up to intermediary energies in the cocoon
where the shear acceleration in sub-relativistic backward flow
occurs.

The source spectra of different components are given by
\begin{equation}
   q(\epsilon ,A)\propto k(A){\epsilon }^{-\gamma }
\exp {\left( -\frac {A\epsilon }{Z\epsilon _{\max }}\right) }
\end{equation}
where the function $k(A)$ describes the source chemical composition and can be written in terms
of the solar composition $k_{\odot }(A)$.

The spectral index $\gamma $, the
maximum energy $\epsilon _{\max }$ and coefficients $k(A)$ adjusted to reproduce observational data are
given in Table 1, see also Section 5 below.

\begin{table}
\begin{center}
\caption{Parameters of source components in Andromeda galaxy}
\begin{tabular}{|c|c|c|c|c|c|}
\hline  &component&$\gamma $&$\epsilon _{\max }$  & $L_{\mathrm{cr}}(z=0)$& $k(A)$\\
\hline  & jet     &  0.5    &$10^{19}$ eV       &$1.3\cdot 10^{40}$ erg s$^{-1}$& 90$k_{\odot }(A),\ A>4$\\
\hline  &bow shock &  2.0    &$5\cdot 10^{15}$ eV&${3.2\cdot }10^{42}$ erg s$^{-1}$&$k_{\odot }(A)A/Z$\\
\hline  &cocoon   &  2.0    &$6\cdot 10^{17}$ eV&${1.4\cdot }10^{41}$ erg s$^{-1}$&$k_{\odot }(A)A/Z$\\
\hline
\end{tabular}
\end{center}
\end{table}

The injection of particles into DSA  depends mainly on the ratio of
atomic mass to the charge. Hybrid modeling of quasi-parallel shocks
shows that the injection is proportional to this quantity
\citep{caprioli17}. Galactic cosmic ray composition then can be
reproduced if ions are injected in the neutral or warm interstellar
medium where they are
 single- or double-ionized. In the case of the hot medium of the galactic halo or the hot cocoon interior,
the full ionization of ions is a more reasonable assumption. So we use a weak enhancement factor of injection $A/Z$ for
  the bow shock and cocoon components. This results in light cosmic ray composition. The use of a similar enhancement
for the cocoon component is justified if the injection occurs in
small-scale shocks in the cocoon or if the bow shock particles are
reaccelerated in the cocoon. The latter opportunity seems very likely
because the halo gas shocked at the bow shock is mixed with
the cocoon material due to development of the Kelvin-Helmholz
instability in the shear flow (see \citet{mbarek19}).

A similar injection mechanism can operate for the jet component when a small number of cocoon particles
can be further reaccelerated in the jet. Since the shear acceleration is very effective and produces the
hardest spectra in ultra-relativistic flows (see e.g. \citet{rieger19,wang21}) we expect that jet particles are mainly
 produced from reacceleration of the bow shock component in the ultra-relativistic part of the jet.
Usually the jet  is ultra-relativistic at sub-kiloparsec scales and
becomes slower at larger distances (see \citet{blandford19} for a review).
Therefore the jet component is produced closer to the jet origin and its
composition is heavy because of the high metallicity of galactic bulges
 and preferential injection of partially ionized heavy ions into DSA at the bow shock.
Qualitatively speaking, the large adjusted enrichment factor 90 of the jet component (see Table 1)
is the product of
the  high metallicity of the galactic bulge 3-5 and the
enhancement factor 10-20 of the ion injection at the bow shock.

\section{Modeling Milky Way and Andromeda galaxy}

\begin{figure}
\begin{center}
\includegraphics[width=8.0cm]{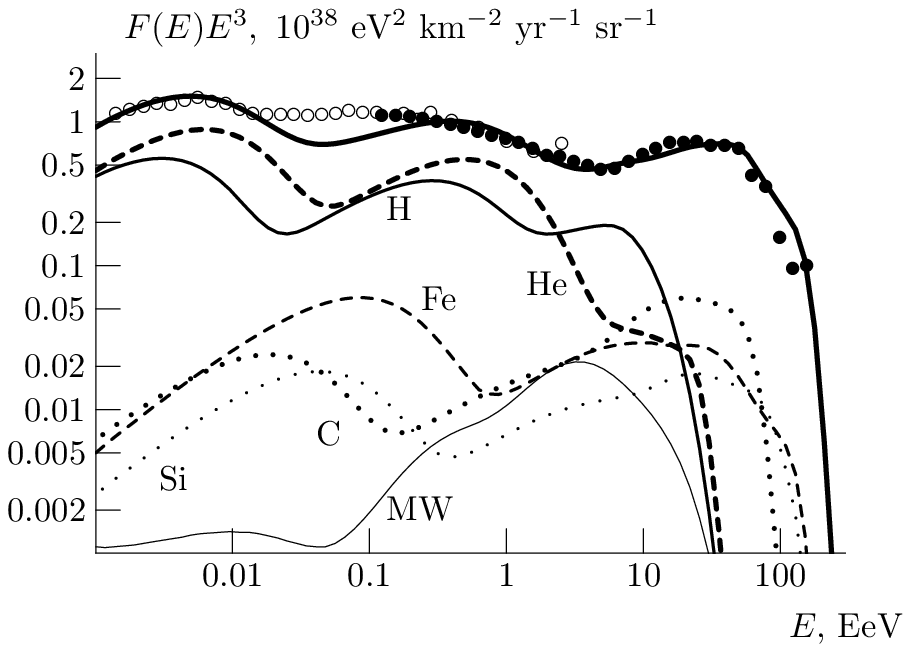}
\end{center}
\caption{ Spectra of different elements and all-particle spectrum (thick solid line) produced in  Andromeda galaxy
 and observed at the Earth position.
A possible contribution in the all particle spectrum from the
Galactic center (MW) is shown by the thin solid line. Spectra of
Tunka-25,  Tunka-133
 array (\citet{budnev20}, open circles) and Pierre Auger Collaboration (\citet{PAO21}, energy shift +10$\% $, black circles) are also shown.}
\end{figure}

\subsection{Andromeda galaxy}
We model the propagation of particles from Andromeda galaxy at distance of 785 kpc from the maximum
redshift $z=1$ down to the present time $z=0$.
It was assumed that the enhanced SMBH accretion produced jet every 280 million years ($\Delta z=0.02$)
with the last episode 140 million years ago ($z=0.01$). In addition we multiply the source terms in Eqs. (1,2) by $(1+z)^4$
to take into account cosmological evolution. Averaged in time luminosity $L_{\mathrm{cr}}$ of different
components of accelerated
particles are given in Table 1.
We use the value of the intergalactic magnetic field
$B=10^{-7}$ G and the coherence scale $l_c=0.13$ Mpc that gives $E_c=1.2\cdot 10^{19}Z$ eV.
The particle distribution vanishes at the spherical boundary
 of simulation domain with radius $R=8$ Mpc that corresponds to the escape of
particles from the edge of the Local Supercluster of galaxies.

The numerical solution of cosmic-ray transport Eqs (1,2) follows the
finite difference method. The results of calculations are shown in
Figures (2-4).

The calculated all-particle spectrum and spectra of protons and
nuclei are shown in Figure 2. The photodisintegration of nuclei on
the background microwave photons strongly influences the high energy
part of the all-particle spectrum. Taking into account the
simplicity of the model the agreement with observations is good. A
slightly
 heavier composition of the bow shock component  would improve the fit at energies $10^{16}-10^{17}$ eV.
The same is true for the calculated mean logarithm A as shown in Figure
3. The slightly heavier  composition is indeed expected because a
small fraction of the time the bow shock propagates in the galactic
bulge where gas is not fully ionized and the gas metallicity is
high. Some input of reacceleration of heavy galactic cosmic ray
component \citep{caprioli15, kimura18b} is not excluded either.

The calculated anisotropy is shown in Figure 4. It is low even despite
 the  large free path of particles that is
 comparable to Andromeda distance for energies above $10^{19}Z$ eV. However the particles traveled to multi-Mpc
distances during the time from the last SMBH activity and this explains the almost isotropic distribution at present.

\begin{figure}
\begin{center}
\includegraphics[width=8.0cm]{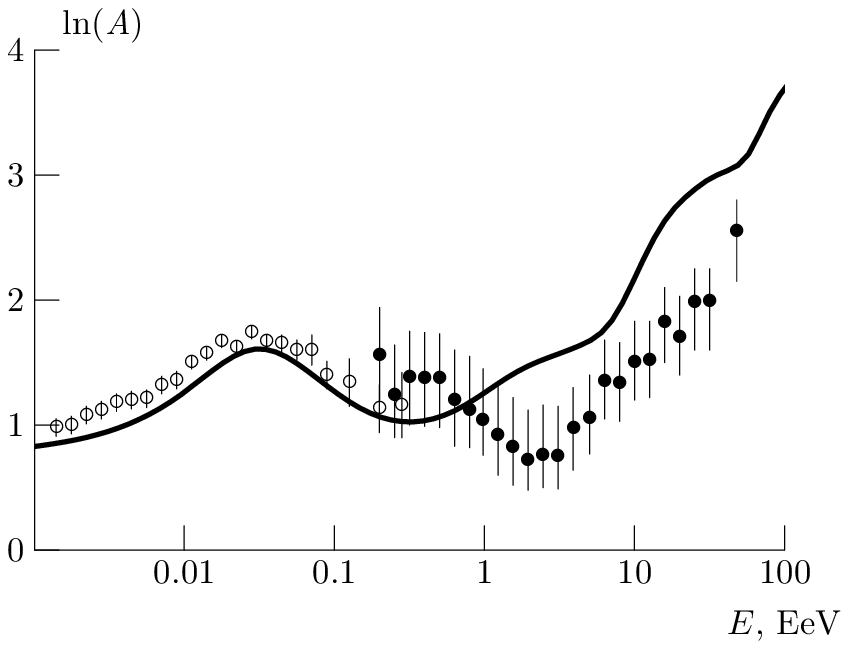}
\end{center}
\caption{ Calculated mean logarithm of atomic number A (solid line). The
measurements of Tunka-133, TAIGA-HiSCORE
 array (\citet{prosin22} open circles) and Pierre Auger Collaboration
(EPOS-LHC, energy shift +10$\% $ \citet{bellido17}, black circles) are also shown.}
\end{figure}

\begin{figure}
\begin{center}
\includegraphics[width=8.0cm]{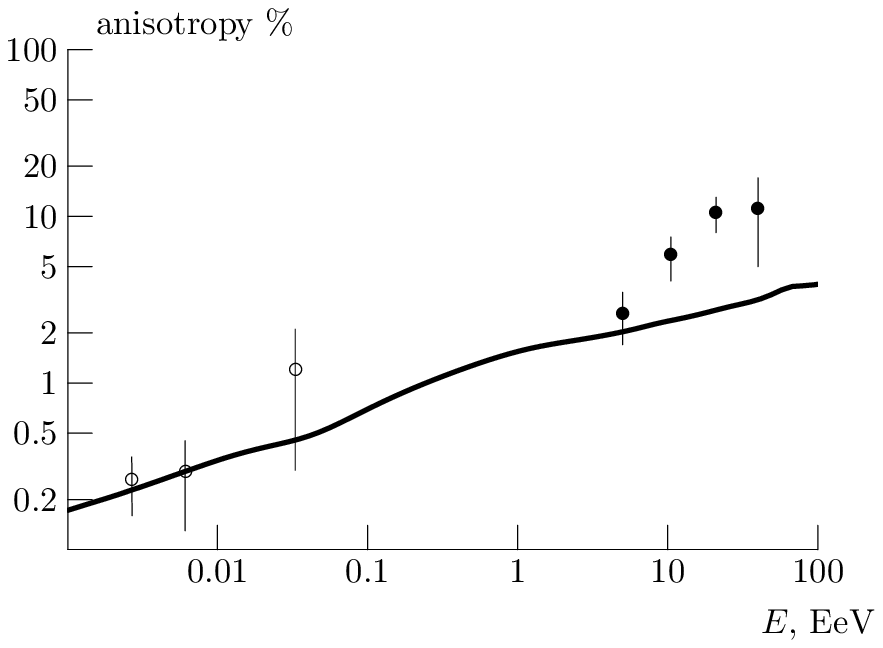}
\end{center}
\caption{ Calculated cosmic ray anisotropy (solid line). The results of Pierre Auger Collaboration
(energy shift +10$\% $, \citet{aab18} black circles)  and KASCADE-Grande
experiment (\citet{chiavassa15} open circles) are also shown.}
\end{figure}

\subsection{Milky Way}
We also model the propagation from SMBH in the Galactic center that
was treated as the scaled Andromeda case. Assuming that the jet power is proportional
to the accretion rate that in turn is proportional to the square of the SMBH mass \citep{bondi52}
 we divide the power
$L_{\mathrm{cr}}$  of the jet, bow shock, and  cocoon  components by $50^2$. The corresponding maximum energies were
 divided by $50$. In addition, we use the last episode of the Galactic SMBH activity
28 million years ago ($z=0.002$) which roughly corresponds to the age of the eROSITA bubbles \citep{predehl20}.
The corresponding all-particle spectrum is shown in Figure 2.

\section{Discussion}

The maximum energy of particles accelerated at the nonrelativistic bow shock is determined
by the nonresonant cosmic ray streaming instability \citep{bell04} (see Appendix for details)

\[
\epsilon ^b_{\max }=\frac {\eta _{\mathrm{esc}}}{2\ln (B/B_b)}e\sqrt{\beta _{\mathrm{head}}L_{\mathrm{j}}c^{-1}}
\]
\begin{equation}
=1.73\cdot {10^{19}} \mathrm{eV} \ \frac {\eta _{\mathrm{esc}}}{2\ln (B/B_b)} \beta ^{1/2}_{\mathrm{head}} \left(
\frac {L_\mathrm{j}}{10^{44}\mathrm{erg}\ \mathrm{s}^{-1}} \right)
^{1/2}
\end{equation}
Here $\beta _{\mathrm{head}}$ is the ratio of the speed of bow hock
"head" to the speed of light $c$,
 $L_{\mathrm{j}}$ is the total power of two opposite directed jets, $\eta _{\mathrm{esc}}$ is the ratio of the energy
 flux of runaway accelerated particles to the kinetic flux  of the shock.
The logarithmic factor in the denominator corresponds to the situation when the seed magnetic field $B_b$ amplified in the
 upstream region of the shock up to values of $B$ via cosmic ray streaming instability.

The parameter $\eta _{\mathrm{esc}}$ is close to 0.01 for shocks where the pressure of accelerated particles
is of the order of 0.1 of the shock ram pressure and can be higher at cosmic-ray modified shocks.
For $\eta _{\mathrm{esc}}=0.01$, $\beta _{\mathrm{head}}=0.1$ and $\ln (B/B_b)=5$ the protons are
accelerated up to multi PeV energies at the jet bow shocks.

Our modeling shows that particles with energies above $10^{15}$ eV can be extragalactic. Lower energy particles are
 probably produced in Galactic supernova remnants.

We found the mean jet cosmic ray power of the order $3\cdot 10^{42}$ erg s$^{-1}$ at the present epoch in the Andromeda galaxy.
Taking into account the standard 10 $\% $ efficiency of DSA we expect the total mean jet power
$3\cdot 10^{43}$ erg s$^{-1}$. Then for 10$\% $ duty cycle the real jet power produced in an active
state will be of the order of
$3\cdot 10^{44}$ erg s$^{-1}$ which is one percent of the Eddington luminosity.
The peak luminosity  could be correspondingly higher for shorter duty cycle of the order of 1$\% $ \citep{bird08}.
Multi-PeV particles indeed can be produced at jet bow shocks in Andromeda galaxy. It takes cosmological time
for these particles to reach the Earth. In this regard pulsations of the source do not play a role at
these low energies. This is opposite to the highest energy part of the spectrum which is mainly determined by the last
 episode of SMBH activity.

The remnants of the cocoon and bow shock produced during the last episode are still in Andromeda's halo at hundred kpc
distances (see, for example, the recent modeling of \citet{husko22}). The electrons are accelerated up to TeV energies at
 the bow shock with the speed of the order of $10^3$ km s$^{-1}$ and produce gamma rays via  Compton scattering of
microwave background photons. The bow shock gamma-ray luminosity is
of the order of the electron bow shock production power. Using  a
proton to electron ratio $10^3$ that is a characteristic value in
young SNRs and the bow shock cosmic ray energetics $3\cdot 10^{42}$
erg s$^{-1}$ mentioned above we obtain the bow shock gamma-ray
luminosity
 $3\cdot 10^{39}$ erg s$^{-1}$ that is in accordance with observations of Andromeda's gamma-ray halo \citep{recchia21}.
Strong electron energy  losses will result in the shell morphology of the gamma-emission
(see also \citet{recchia21} for details).

The scaled jet power in the Galactic center is of the order of  $10^{41}$ erg s$^{-1}$ in the active state that is
 exactly what is needed for the production of eROSITA bubbles \citep{predehl20}.
With such energetics the Galactic center gives a small contribution in observed all particle spectrum
 (see Figure 2). We leave a detail treatment of the Galactic center's contribution
for future investigations.

In is known that the electric potential difference is a reasonable
estimate for the maximum energy of particles accelerated at
quasi-perpendicular shocks \citep{zirakashvili18}. For example,
single charged anomalous cosmic rays are accelerated up to hundreds
MeV at the solar wind termination shock with the electric potential
200 MV {\citep{cummings87}}. The jet electric potential is also a
good estimate for the maximum energy as seen in trajectory
calculations \citep{alves18, mbarek19}.

It is given by
\[
\epsilon ^j_{\max }=e\sqrt{\beta _{\mathrm{j}}L_{\mathrm{mag}}c^{-1}}
\]
\begin{equation}
=1.73\cdot {10^{19}} \mathrm{eV} \ \beta _{\mathrm{j}}^{1/2} \left(
\frac {L_\mathrm{mag}}{10^{44}\mathrm{erg}\ \mathrm{s}^{-1}} \right)
^{1/2}
\end{equation}
where $L_{\mathrm{mag}}$ is the magnetic luminosity of two opposite jets.

So the maximum energy of the jet component $10^{19}$ eV used in our calculation is also in agreement
with theoretical expectations.

Our model does not exclude a contribution at the highest energies from more distant sources like Cen A or M87.
For example, \citet{mollerach19} use similar propagation parameters and showed that UHECRs
can originate in Cen A.

It is known that the dipole anisotropy of the Auger Collaboration is
in the direction of Cen A \citep{aab18}. Note that the direction of
the anisotropy does not necessarily coincide to the direction of the
main source. It could be that the particles produced during the last
event in Andromeda are distributed isotropically within several Mpc
now while the currently active Cen A source produces the observed
anisotropy. In addition, there are several other "hot spots". In
particular, the Telescope Array Collaboration reported the detection
of the "hot spot" in the direction of the Perseus-Pisces
supercluster \citep{kim22}. The Andromeda galaxy is in the same
direction. So
 we can not exclude that this excess is related to Andromeda (see also the recent paper of \citet{plotko22}).

\section{Conclusion}
Our results and conclusions are the following:

1) We performed the modeling of the propagation of UHECRs produced in the Galactic center and in the nearby Andromeda galaxy.
It was assumed that the periodic activity of the central SMBHs produces
large-scale jets accelerating high energy particles.

2) We found that the light intermediary energy component of the jet
cocoon produced via shear acceleration mechanism can explain the
observable spectrum and composition below the "ankle". Heavier higher
energy component with hard spectrum is probably produced in the jet
itself.

3) Lowest energy light component related to particles
accelerated at the bow shock can explain the cosmic ray spectrum at PeV energies.

4) The Andromeda's gamma-ray halo is produced by electrons currently accelerated at the bow shock
propagating in the galactic halo.

5) The production of UHECRs in Andromeda galaxy can explain the "hot
spot" observed by Telescope Array Collaboration \citep{kim22}.

6) We suggest some modification of the reacceleration of galactic cosmic
rays by jets \citep{caprioli15, kimura18b}. It seems that the
reacceleration of bow shock particles makes the main contribution to the production of UHECRs.
The heavy composition is expected because the reacceleration
efficiency is highest for ultra-relativistic jets. This happens
close to the jet origin in the galactic bulge where the gas is
partially ionized and has a high meatallicity.

\section*{Acknowledgements}
We thank the anonymous referee for useful comments.
The work was partly performed at the Unique scientific
installation "Astrophysical Complex of MSU-ISU"
(agreement 13.UNU.21.0007).

\section*{Data Availability}

All results in this paper were obtained using available published data.



\bibliographystyle{mnras}
\bibliography{jet} 

%






\appendix
\section*{APPENDIX. Maximum energy limit from the streaming instability}
The maximum rate $\Gamma $ of the nonresonant streaming instability is given by \citep{bell04}

\begin{equation}
\Gamma =\frac {\sqrt{\pi }J}{c\sqrt{\rho }}=\frac {\eta _{\mathrm{esc}}u^3e\sqrt{\pi \rho}}{2c\epsilon _{\max }}
\end{equation}
where $\rho $ is the plasma density, $u$ is the shock speed, and the electric current of energetic
particles $J$  was expressed in terms of the parameter $\eta _{\mathrm{esc}}$ that is the ratio of the energy flux of runaway
 particles with energy $\epsilon _{\max }$ to the flux of the shock kinetic energy  $\frac 12\rho u^3$.

For the instability to have enough time to amplify the magnetic field from the seed value of $B_b$ to the value of $B$ the rate
$\Gamma =\frac uR\ln{\frac {B}{B_b}}$ where $R$ is the shock radius. This gives the estimate for the maximum energy
(c.f. \citet{zirakashvili08, bell13})

\begin{equation}
\epsilon _{\max } =
\frac {e\eta _{\mathrm{esc}}\sqrt{\pi \rho }Ru^2}{2c\ln{{B}/{B_b}}}
=\frac {\eta _{\mathrm{esc}}}{2\ln (B/B_b)}e\sqrt{uL_{\mathrm{j}}c^{-2}}
\end{equation}
where the shock parameters are expressed in terms of the power of two opposite jets $L_\mathrm{j}=\pi \rho u^3R^2$.
Introducing $\beta =u/c$ we obtain the equation (5) in the main text.


\bsp    
\label{lastpage}
\end{document}